# Optimum stabilization of self-mode-locked quantum dash lasers using dual optical feedback with improved tolerance against phase delay mismatch

HAROON ASGHAR,* EHSAN SOOUDI, PRAMOD KUMAR, WEI WEI AND JOHN. G. MCINERNEY

*Department of Physics and Tyndall National Institute, University College Cork, Cork, Ireland*
*\*haroon.asghar@ucc.ie*

**Abstract:** We experimentally investigate the RF linewidth and timing jitter over a wide range of delay tuning in a self-mode-locked two-section quantum dash lasers emitting at ~ 1.55_m and operating at ~ 21 GHz repetition rate subject to single and dual optical feedback into gain section. Various feedback conditions are investigated and optimum levels determined for narrowest linewidth and reduced timing jitter for both single and dual loop configurations. We demonstrate that dual loop feedback, with the shorter feedback cavity tuned to be fully resonant, followed by fine tuning of the phase of the longer feedback cavity, gives stable narrow RF spectra across the widest delay range, unlike single loop feedback. In addition, for dual loop configurations, under fully resonant conditions, integrated timing jitter is reduced from 3.9 ps to 295 fs [10 kHz-100 MHz], the RF linewidth narrows from 100 kHz to < 1 kHz, with more than 30 dB fundamental side-mode suppression. We show that dual loop optical feedback with separate fine tuning of both external cavities is far superior to single loop feedback, with increased system tolerance against phase delay mismatch, making it a robust and cost-effective technique for developing practical, reliable and low-noise mode-locked lasers, optoelectronic oscillators and
pulsed photonic circuits.

## 1. Introduction

Quantum confined, passively mode-locked diode lasers (PMLLs) have attracted much attention in recent years due to increasing interest in highly stable ultrafast pulse trains with low timing jitter, and high repetition rates (tens of GHz), producing a comb of phase-locked modes spaced at the repetition frequency [1]. The phase noise of modal beat note in optical frequency combs should be optimized for various applications in optical communication systems, such as multi-carrier transmission systems in orthogonal frequency division multiplexing (OFDM) [2], coherent wavelength division multiplexing (CoWDM) [3], arbitrary waveform generation [4], all optical signal processing [5] and millimeter-wave generation [6]. To improve the phase noise of PMLLs, several experimental methods such as external optical feedback [7-11], coupled optoelectronic oscillators (OEOs) [12-14], hybrid mode-locking [15] and injection locking [16, 17] have been proposed and demonstrated. Optoelectronic feedback has been utilized to stabilize timing jitter by conversion of the optical signal (using a fast photodetector) to an electrical oscillation for use in a long feedback loop. This technique does not utilise an RF source, but requires optical-to-electrical conversion. Hybrid mode-locking requires electrical modulation of the gain or absorber bias, while optical injection needs an external laser, making these techniques less attractive for practical implementation where low cost, simplicity and reliability are important. Of all stabilization techniques demonstrated to date, optical feedback with long delay is the most practical approach to reduce the phase noise of PMLLs, but side-mode spurs in the signal spectrum are still a major problem. These noise-induced side-bands contribute significantly to timing jitter, particularly [11]. Recently, a simpler technique using a second loop shorter than the main one, dual optical feedback [18], has been shown to improve timing jitter of the PMLLs and to suppress external cavity side-modes. This dual loop feedback configuration gives sub-kHz linewidth with 38 dB suppression in the fundamental side-mode. A number of experimental [18-21] and numerical investigations [22] have also been made of the dynamics of mode-locked semiconductor lasers subject to dual loop feedback. We note that optimum performance of RF linewidth and timing jitter in dual loop feedback has been very sensitive to small phase adjustments, which limits operation to a narrow parameter space and makes optimum performance vulnerable to environmental change. For practical applications of MLLs, it is desirable to reduce the sensitivity of resonant feedback on the feedback delay phase, such that changing delay length maintains stable RF spectra with narrow RF linewidth on the wide delay range. To the best of our knowledge, this important requirement has not been addressed in the literature. In this paper, we thoroughly investigate the experimental parameter range for delay times for a two-section quantum dash mode-locked laser in a dual loop feedback scheme. We also demonstrate optimum RF linewidth, and timing jitter for wide range of delay tuning making this scheme ideal for practical applications. Furthermore, the effect of feedback ratios for single and dual loop feedback on the noise properties of the mode-locked laser are discussed. Setting the length of shorter feedback cavity to the fully resonant value, then phase tuning of the longer feedback cavity results in narrower RF linewidth and lower timing jitter across a wide delay range, compared to single loop feedback. Moreover, it was further noticed that for optimized dual loops, more than 30 dB suppression in the fundamental side-mode is obtained due to the interference of the two delayed optical signals at the 3-dB coupler. Under stably resonant conditions, optimal feedback level (~ -22 dB) and constant temperature control ($19^0$C), the RF linewidth is reduced from 100 kHz for the free running case to < 1 kHz (instrument limited) for dual loop and 4 kHz for single loop feedback. RMS timing jitter (integrated from 10 kHz-100 MHz) is also reduced from 3.9 ps free running to 295 fs for dual loops and 700 fs for single loop feedback. A global study of single and dual loop setups reveals that controlled dual loop feedback, with precise alignment of phase delays in both external cavities, maximizes the range of effective feedback conditions compared to single loop feedback, with minimal additional complexity and cost.

## 2. Experimental Setup

The device under investigation is a two-section InAs/InP quantum dash MLL with an active region consisting of nine InAs quantum dash monolayers grown by gas source molecular beam epitaxy (GSMBE) embedded within two barrier layers (dash-in-barrier device), and separate confinement heterostructure layers of InGaAsP, emitting at ~ 1.55 μm [23]. Cavity length was ~ 2030 μm, 11.8% (240 μm) of which formed the absorber section, giving pulsed repetition frequency ~ 20.7 GHz ($I_{Gain}$ = 300 mA,) and average power of ~ 0.7 mW in fiber, per facet. Mode-locking was obtained without reverse bias applied to the absorber section. Heat sink temperature was fixed at $19^0$C. This is a two-section device but is packaged similarly to a single section self-mode-locked laser since the absorber is not biased: its minimal absorption does not affect the self-mode-locking mechanism [16]. Gain and absorber layers were electrically isolated by 9 kΩ. The MLL was mounted p-side up on an AlN submount and a copper block with active temperature control. Electrical contacts were formed by wire bonding.

{A schematic of our feedback experiment is depicted in Fig. 1. For single and dual loop feedback, a calibrated fraction of light was fed back through port 1 of an optical circulator, then injected into the laser cavity via port 2. Optical coupling loss from port 2 to port 3 was -0.64 dB. The output of the circulator was sent to a semiconductor optical amplifier (SOA) with a gain of ~ 9.8 dB, then split into two arms by a 50/50 coupler. 50% went to an RF spectrum analyzer (Keysight E-series, E4407B) via a 21 GHz photodiode and optical spectrum analyzers (Ando AQ6317B and Advantest Q8384). The other 50% of power was directed to the feedback operations. For a single feedback loop, all power passed through loop-I. For dual loop configurations (feedback loops-I and-II) the power was split into two equal parts via 3-dB splitter. Each feedback loop contained an optical delay line combined with a variable optical attenuator and polarization controller. Loop lengths were 160 m and 80 m corresponding to pulse round-trip frequencies 1.28 MHz and 2.60 MHz, respectively. Feedback strengths in both loops were controlled by variable optical attenuators and monitored using the power meter. Equal feedback was received from both external feedback cavities. The microscopic lengths of the fiber loops were optimized by optical delay lines based on stepper-controlled stages with resolution 1.67 ps. Polarization controllers in each loop plus a polarization controller before port 1 of the circulator ensured the light fed back through both loops matched the emitted light polarizations to maximize feedback effectiveness.

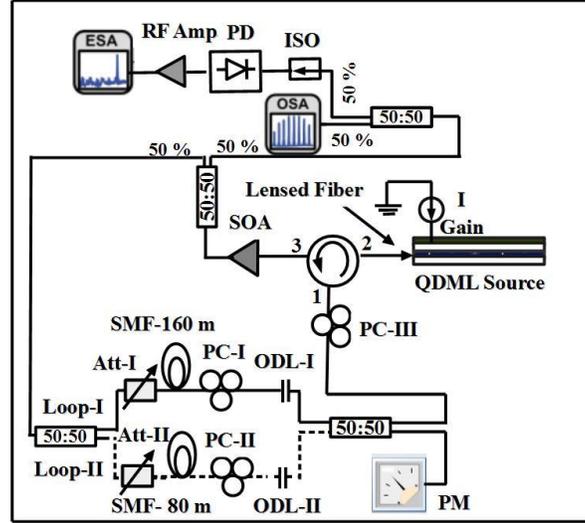

Fig. 1. Schematic of the experimental arrangement for single (excluding dashed portion) and dual loop configurations (with dashed portion). *Acronyms*—SOA: Semiconductor Optical Amplifier; ISO: Optical isolator; PD: Photodiode; ODL: Optical delay line; Att: Optical attenuator; PC: Polarization controller; ESA: Electrical spectral analyzer; OSA: Optical spectrum analyzer; SMF: Single mode fiber; PM: Power Meter

In this work, the RMS timing jitter $\sigma_{RMS}$ is calculated from the single sideband (SSB) phase noise spectra measured for the fundamental RF frequency (≈ 20.7 GHz) using [10]:

$$\sigma_{RMS} = \frac{1}{2\pi f_{ML}} \sqrt{2\int_{f_d}^{f_u} L(f) df}$$

Where $f_{ML}$ is the pulse repetition rate and $f_u$ and $f_d$ are the upper and lower integration limits. L(f) is the single sideband (SSB) phase noise spectrum, normalized to the carrier power per Hz. To measure RMS timing jitter of the laser in more detail, single-sideband (SSB) noise spectra for the fundamental harmonic repetition frequency were measured. To assess this, RF spectra at several spans around the repetition frequency were measured from small (finest) to large (coarse) resolution bandwidths. The corresponding ranges for frequency offsets were then extracted from each spectrum and superimposed to obtain SSB spectra normalized for power and per unit of frequency bandwidth. The higher frequency bound was set to 100 MHz (instrument limited).

To investigate the effects of external optical feedback on RF linewidth and integrated timing jitter in single and dual loop configurations, the attenuation in the gain section was varied from the minimum achievable feedback level ~ -46 dB to the maximum feedback before the laser became unstable ~ -22 dB. These results demonstrate that at ~ -46 dB, the RF linewidth is 80 kHz for single and 70 kHz for dual loop feedback, with corresponding timing jitter 3 ps and 2.75 ps, respectively, [10 kHz-100 MHz]. At this low attenuation, the effects of external optical feedback are very small, so that no major reduction in RF linewidth and timing jitter were seen relative to free running. However, with increased feedback ratio to ~ -29 dB, a gradual decrease in the RF linewidth and RMS jitter were observed. This feedback level (~ -29 dB) gives RF linewidth as low as 12 kHz for single loop and 10 kHz for dual loops, and jitter reduced to 1.4 ps and 1.38 ps, respectively. Further increase to ~ -22 dB results in saturated RF linewidth and timing jitter. The minimum achieved RF linewidth and timing jitter as functions of feedback ratio for integer resonant cases are depicted in Fig. 2.

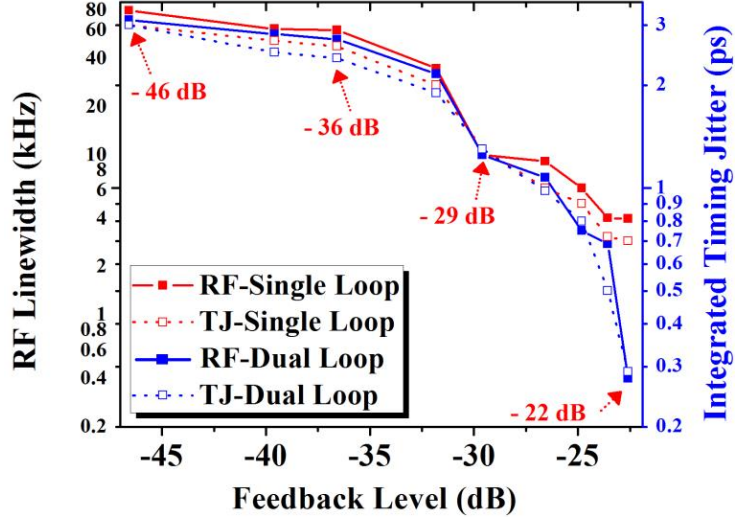

Fig. 2. 3-dB RF linewidth and integrated timing jitter as a function of external feedback ratio at a bias of 300 mA gain current for single and dual loop feedback configurations

From this detailed analysis of RF linewidth versus feedback strength, we have identified optimal feedback parameters for single and dual loop configurations, with strong reduction in the RF linewidth and lowest timing jitter of these MLLs. Our results demonstrate that for practical applications, the relatively flat characteristics of RF linewidth versus feedback ratio [-24 dB, -23 dB and -22 dB] are more favorable. Variation in RF linewidth and timing jitter in both schemes follows a similar trend when feedback approaches the optimal value, which agrees well with reported analytical expression (square root dependence of RF linewidth on integrated timing jitter) [24]. Recently, for a quantum dot mode-locked laser operating at ~ 5.1 GHz, the minimum RF linewidth was obtained at relatively low feedback level -36 dB [8]. On the other hand, it is noted for a passively mode-locked quantum dash laser emitting at 1580 nm and operating at 17 GHz repetition rate, marked reduction in RF linewidth occurs at significantly stronger feedback -22 dB [9], in agreement with our studies. These differences are explicable by the likelihood that the antiguiding (phase-amplitude coupling) factor is lower in quantum dashes.

To study the dynamic effects of single loop feedback on RF linewidth and noise properties of MLLs, loop-II was disconnected and maximum feedback to the gain section was limited to ~ -22 dB. To achieve stable resonant conditions for the single loop, the microscopic length of the feedback cavity was optimized using an optical delay line (ODL-I) adjustable from 0 to 84 ps in steps of 1.67 ps. This optimization of the single feedback loop delay reduced the RF linewidth considerably, as for other reported experiments [7] and theoretical predictions [25]. Resulting RF linewidth (black squares) and timing jitter (blue triangles) are shown in Fig. 3 versus delay tuned from 0-84 ps; clearly the efficacy of single loop feedback is highly dependent on feedback delay.

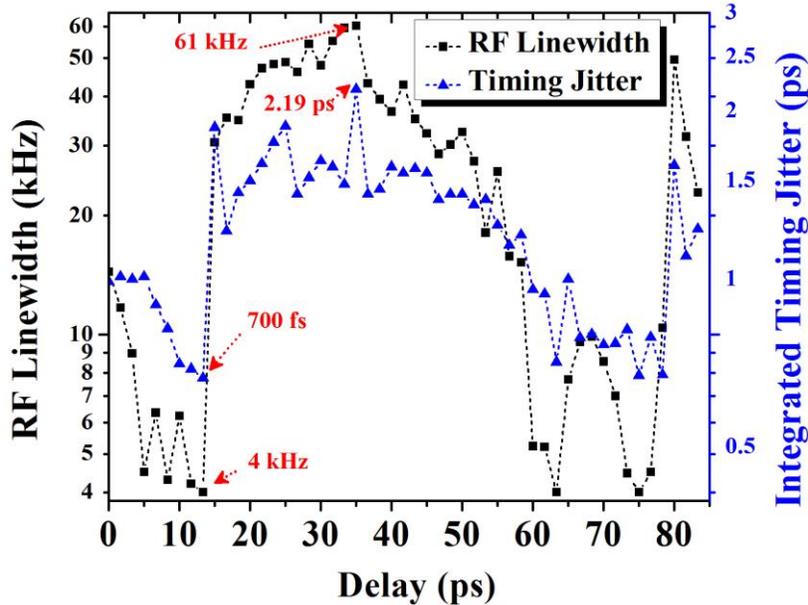

Fig. 3. RF linewidth (black squares) and integrated timing jitter data (blue triangles) for mode-locked pulse trains as functions of delay phase [0 – 84 ps] for single optical feedback

It can be seen from the ranges 0-13.4 and 60-78.4 ps that effective stabilization is achieved, when the external cavity length is close to an integer multiple of that of the solitary laser. When fully resonant (13.4 ps and 63.4 ps), the RF linewidth decreased from 100 kHz free running to 4 kHz, with SSB phase noise - 70 dBc/Hz at frequency offset 10 kHz. Due to this decrease in phase noise, the timing jitter reduced from 3.9 ps free running to ∼ 700 fs [10 kHz - 100 MHz]. Lorentzian fit of the RF linewidth under 1 MHz frequency span with resolution bandwidth 1 kHz and video bandwidth 100 Hz is shown in the inset of Fig. 4 (a). Measured phase noise traces for free running laser (black line) and single loop feedback (blue line) as functions of frequency offset from fundamental mode-locked frequency are depicted in Fig. 4 (b). When fully resonant condition (13.4 ps), the peak power of the RF spectrum rises by -52 dB. However, for the free running condition the peak power of RF spectra is observed to be -20 dB. For single loop feedback, -32 dB increase in peak power of RF spectra is a result of reduced RF linewidth and enhanced threshold current which contributes to increase in optical power. The peak power of the RF spectrum as a function of full delay phase tuning 0 - 84 ps is presented in Fig. 4 (a). In addition, we observed that, at delays from 15-58.4 ps and 80-83.4 ps, the laser became highly unstable in its noise emission: RF linewidth broadened to 61 kHz and RMS timing jitter to 2.2 ps at delay 35 ps. Experimental results on single loop feedback show that for practical use of MLLs, the most suitable and stable delay ranges are close to resonant regimes (0-13.4 ps and 60-78.4 ps). However, this is still quite sensitive in packaging and production. In the next section we describe how to reduce the sensitivity of resonant single loop feedback using a stable and efficient dual loop feedback scheme which is practical, robust and cost-effective.

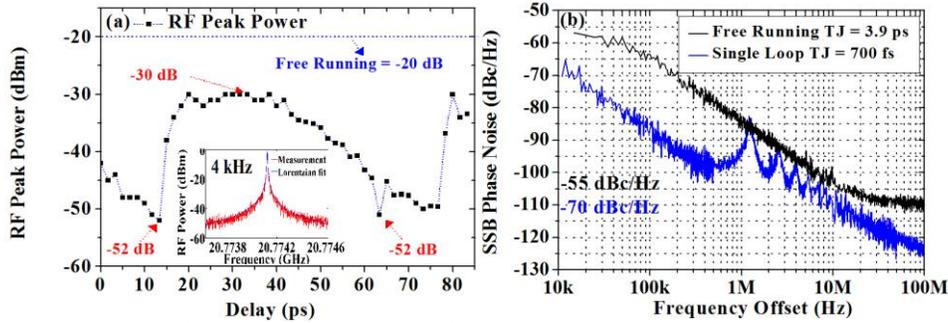

Fig. 4. (a) Peak power of RF spectrum as a function of full delay tuning [0 – 84 ps]; Inset shows measured RF spectra and Lorentzian fit for single loop (b) Comparison of SSB phase noise traces of single loop feedback (blue line) with free running condition (black line)

In dual loop feedback, simultaneous optical feedback from two external cavities are applied to the gain section at ~ -22 dB. The feedback from both cavities is kept equal using variable optical attenuators (Att-I and Att-II) plus fine adjustment of polarization controllers (PC-I and PC-II) in both feedback loops. The optical delay (ODL-II) was adjusted to full resonance and the length of optical delay line (ODL-I) was tuned from 0-84 ps, the maximum range available. This arrangement yielded much better dynamics: stable narrow RF spectra were maintained across the full delay range 0 - 84 ps, unlike single loop feedback. Measured RF linewidth (black squares) and timing jitter data (blue triangles) are shown in Fig. 5.

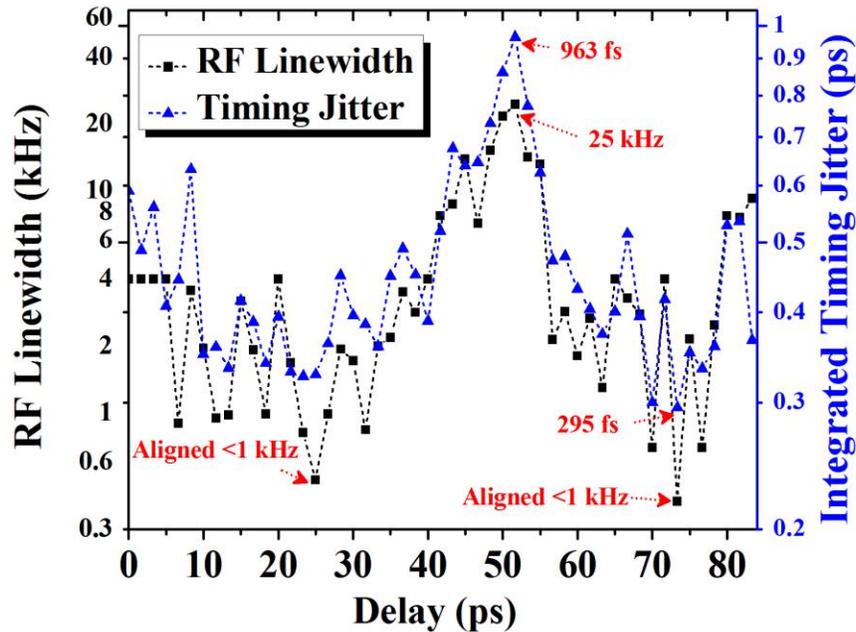

Fig. 5. RF linewidth (black squares) and integrated timing jitters (blue triangles) of mode-locked pulse trains as a function of full optical delay tuning in dual loop feedback

It was observed that at multiple optical delay ranges [Fig. 6 (a)] the RF linewidth was instrument limited at less than 1 kHz, so the actual value may be lower. Lorentzian fit of RF linewidth under 1 MHz frequency span with resolution bandwidth 1 kHz and video bandwidth 100 Hz is shown in inset of Fig. 6 (a). This behavior indicates that to maximize the RF linewidth, delicate tuning of both optical delay lines is required. Furthermore, in this experimental arrangement, from delay settings 0-40 ps and 56.7-78.4 ps, the RF linewidth was below the minimum RF linewidth measured for single loop feedback; refer to Fig. 6 (b). Our results demonstrate that dual loop feedback is more effective than single loop feedback at reducing linewidth and jitter, across a much wider range of delay phase. The resonant condition in dual loop feedback is nearly independent of optical delay, making it ideal for practical applications where robustness and tolerance to misalignment are essential.

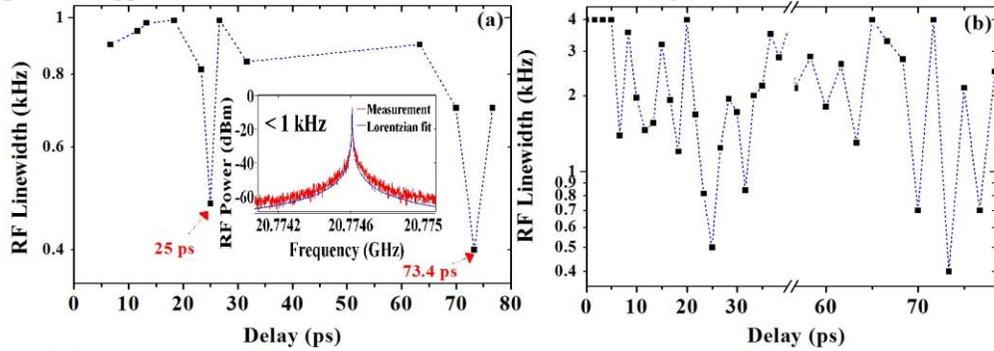

Fig. 6. RF linewidth as a function of optical delay (a) less than 1 kHz (instrumental limited); Inset shows measured RF spectrum and its Lorentzian fit (b) RF linewidth as a function of optical delay [0-40 ps and 56.7 - 78.4 ps] less than 4 kHz (minimum RF linewidth measured in single loop feedback)

We also assessed the influence of dual loop optical feedback on side-mode suppression ratio (SMSR). When optical delay line ODL-I was retuned to 25 ps and ODL-II to 15 ps, every second mode of loop-I coincided precisely with a mode of loop-II. As a result, a maximum of 30 dB suppression in the first order side-mode occurs [See Fig. 7 (c)] and RF linewidth narrowed to 1 kHz, with a SSB phase noise -80 dBc/Hz at frequency offset 10 kHz; timing jitter was reduced to 295 fs. Optimal suppression of external cavity side-modes occurred due to interference of the two delayed feedback signals at the 3-dB coupler. Modal overlap in the RF spectrum was observed for spectral alignment of the modes of the two fiber loops having identical intensity, forming an effective RF frequency comb [See Fig. 7 (c)]. Exact conditions for modal overlap, of course, depend on the precise optical lengths of the additional shorter feedback loop; here we saw 2.60 MHz spacing between supermodes, consistent with our ~ 80 m outer loop. However, with a slight tuning offset between the modes of both fiber loops, strong side-modes appeared with intensity -26 dB [See Fig. 7 (d)], at the same time RF linewidth was reduced to 1.94 kHz and integrated timing jitter to 450 fs. The SSB phase noise traces for dual loop subjected to aligned (blue line) and misaligned (black line) cavity states are depicted in Fig. 8 (a).

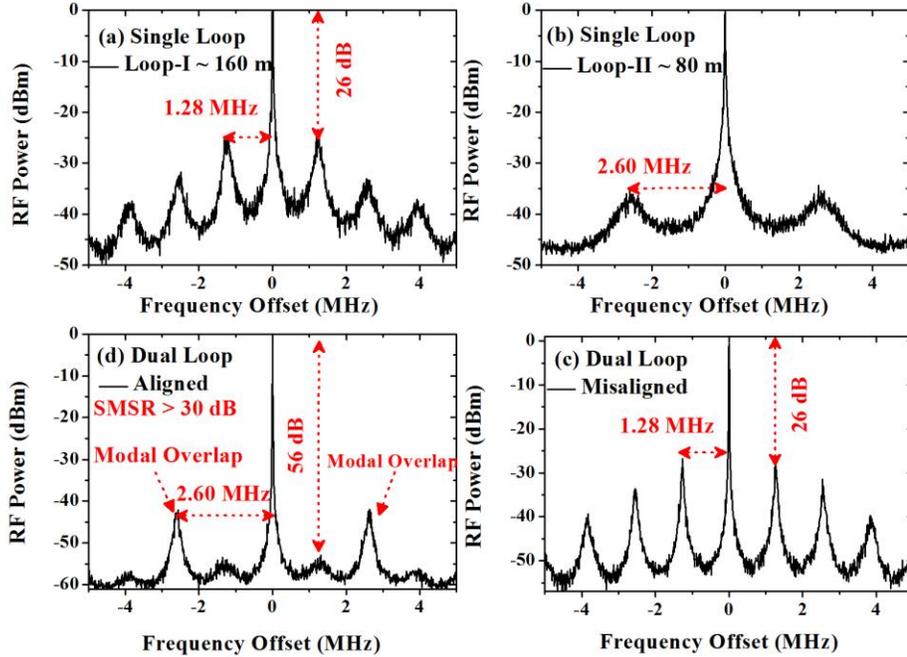

Fig. 7. RF Spectra of single loop feedback with length (a) 160 m (b) 80 m (c) dual loop having spectrally aligned cavity lengths with > 30 dB sidemode suppression (d) RF spectrum of spectrally offset (misaligned) dual loop cavity with strong side-mode: All spectrums are measured using Span=10 MHz, resolution bandwidth=10 kHz and video bandwidth = 1 kHz

From measured RF spectra of single optical feedback and misaligned dual loop feedback, the expected noise resonances were seen near the fundamental frequency of the laser, contributing significantly to the phase noise. The 155 fs reduction in timing jitter for aligned cavity state as compared to misaligned further confirms this argument. However, dual loop feedback with precise adjustment of delay in both loops, and interference of the optical signals at the 3-dB coupler effectively reduce the timing jitter due to suppression of unwanted noise-induced oscillations. Moreover, from the RF spectra, it is noticeable that the peak power of the RF spectrum in dual loop spectra rises by -63 dB. The varying behavior in RF peak power as a function of full delay phase tuning [0 - 84 ps] is shown in Fig. 8 (b).

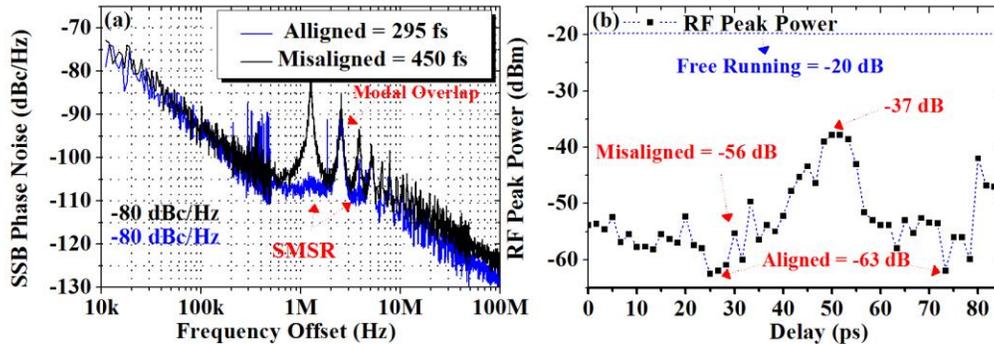

Fig. 8. (a) SSB phase noise traces for misaligned dual loop (black line) and aligned dual loop (blue line) with integration limits 10 kHz - 100 MHz (b) Variation of peak power of RF spectra as a function of full delay tuning [0-84 ps]

## 3. Conclusion

We have reported the effects of optical feedback from dual and single feedback loops on the RF linewidth and timing jitter for wide range of delay tuning in self-mode-locked two-section quantum dash lasers emitting at ~ 1.55 µm and operating at ~ 21 GHz repetition rate. Mode-locking occurred without reverse bias applied to the absorber section, which greatly simplifies packaging and optimization compared to reverse biased systems. Our data reveal that dual loop feedback maximizes tolerance to delay phase mismatch relative to single loop feedback. Moreover, optimized dual loop feedback extends the effective resonant feedback regime and maintains stable RF spectra, with narrow RF linewidth and reduced timing jitter across the entire accessible delay range, making this setup desirable for practical applications. Furthermore, under fully resonant conditions, RF linewidth narrows from 100 kHz free running to below the resolution bandwidth of our ESA (1 kHz) for dual loops and 4 kHz for single loop feedback. Similarly, the integrated timing jitter is reduced from 3.9 ps free running to 295 fs for dual loop and 700 fs for single loop feedback. In addition, when both feedback cavities are fully resonant, external cavity side-modes are suppressed by more than 30 dB relative to single loop feedback. These results suggest that optimized dual loop feedback is a robust and effective means to overcome the most difficult performance limitations of mode-locked diode lasers, namely large timing jitter and tendency to exhibit instabilities. This could allow their major advantages - compactness, ruggedness, a wide range of available wavelengths, low voltage operation, ability to sustain very high repetition rates, mass manufacturing techniques - could be exploited in future optical comb and terahertz generation schemes and stable photonic oscillators.

## Acknowledgments

The authors acknowledge financial support from Science Foundation Ireland (grant 12/IP/1658) and the European Office of Aerospace Research and Development (grant FA9550-14-1-0204).